\documentclass{article}

\usepackage[final]{DLI_2023}

\usepackage[utf8]{inputenc} % allow utf-8 input
\usepackage[T1]{fontenc}    % use 8-bit T1 fonts
\usepackage{hyperref}       % hyperlinks
\usepackage{url}            % simple URL typesetting
\usepackage{booktabs}       % professional-quality tables
\usepackage{amsfonts}       % blackboard math symbols
\usepackage{nicefrac}       % compact symbols for 1/2, etc.
\usepackage{microtype}      % microtypography
\usepackage{xcolor}         % colors

\usepackage[
backend=biber,
sorting=ynt
]{biblatex}

\title{The Glamorisation of Unpaid Labour: \\ AI and its Influencers 
}

\author{%
  Nana Mgbechikwere Nwachukwu \\
  Berkman Klein Center for Internet and Society at Harvard \\
  \texttt{mgn721@harvard.edu} \\
  \And
 Jennafer Shae Roberts \\
  Accel AI Institute \\
 \texttt{jennafershae@accel.ai} \\
  \AND
  Laura N Montoya \\
  Accel AI Institute \\
  \texttt{laura@accel.ai} \\
}

\begin{document}

\maketitle

\begin{abstract}
 To harness the true potential of Artificial Intelligence (AI) for societal betterment, we need to move away from prioritising corporate interests which exploit Global South workers in the digital age.  The unpaid labour and societal harms which are generated by Digital Value Networks (DVNs) disproportionately affect workers in Africa, Latin America, and India and need to be regulated. In this research, we discuss unethical practices to automate Human Intelligence Tasks (HITs) through gig work platforms and the capitalisation of data collection utilising influencers in social media. These are important areas of study in worker and user data practices, where ethical AI could be impactful. We provide suggestions for a path forward focused on responsible AI development. 

\end{abstract}

\section{Introduction }

Microtask marketplaces stimulate an environment where companies and academics may outsource tedious jobs for AI training while paying less to data workers who earn just enough to justify the flexibility and convenience of the platform.\cite{Hitlin2016ResearchStudy}\cite{Moss2023IsWages} Human Intelligence Tasks (HITs) may include work that is deemed too complex for computers, or where algorithms are still prone to low accuracy. Common HITs for AI training include data scraping, data labeling, content moderation, transcription, data generation, survey completion, and data entry.\cite{Ross2010WhoCrowdworkers} Scientists are questioning whether it is ethical to continue the use of gig-pay microtask marketplaces for crowdsourced data collection and processing with consideration of the workers' exploitation, \cite{Gebru2021ForTech}\cite{Damer2019StopResearch}\cite{Newman2019IHour.}  low wages offered per task,\cite{Hitlin2016ResearchStudy} and exposure to stressful or potentially hazardous environments.\cite{Moss2023IsWages} While scholars in the U.S. must get approval from an Institutional Review Board (IRB) to conduct research involving human subjects, including crowd workers, companies hiring on these platforms and those academics outside the US are not held to such standards.\cite{Xia2023TwoMTurk} The IRB follows the Bellmont Principles, a set of ethical guidelines for use in research in order to ensure the protection of human subjects. The Bellmont principles for research involving human subjects include respect for persons, beneficence, and justice.\cite{DepartmentofHealth2014TheResearch.} These principles are not recognized during Business Process Outsourcing (BPO) in other countries. For example, BPO by French Startups for HIT labour in Madagascar to enhance AI capabilities through data labelling and annotation allows for long-term crowdsourced subcontracting with little regard for workers' rights.\cite{LeLudec2023TheMadagascar}

\paragraph{}
Amazon Mechanical Turk, launched in 1995 is estimated to be the largest global source for outsourced work with over half a million registered participants.\cite{Pittman2016AmazonsResearch} Workers are often paid pennies per task, resulting in earnings far below the minimum wage in many developed countries.\cite{Hitlin2016ResearchStudy}  However, this rate can still be higher than average wages in some parts of the Global South, leading to a form of economic exploitation. Task valuation is often set by the task requester, leading to significant pay discrepancies. Tasks that require a high level of skill or a large time investment may not pay proportionally. Pay rates can vary significantly from task to task and requester to requester, with no standard pay rate or policy. Workers have limited rights and protections, and have little to no recourse in the event of disputes. A recent survey of 2,026 United States Mechanical Turk workers whose demographics and education level mirrored the US population, found that it is possible to earn \$30.05 per hour working full-time on the platform, with the medium being \$10.26 per hour.\cite{Hitlin2016ResearchStudy} While these platforms provide valuable income opportunities for many individuals in Africa, Latin America, and India, the broader economic implications, including the potential for exploitation and perpetuating cycles of poverty, are significant and problematic.

\paragraph{}
The gains for companies that employ through these Digital Value Networks (DVNs) often include unpaid labour in the form of time spent by crowd-workers to maintain access and reputation in the marketplace, time spent applying for micro-tasks, improving skills through certification, and non-payment for completed work that is denied by the client with no repercussions.\cite{Howson2022UnpaidNetworks} DVNs disproportionately bias crowd-workers in the Global North due to geographic location, burdening those in the Global South with more unpaid labour. Unpaid labour is also common in marketing by companies through the use of social media influencers. Often neither the influencer nor those being influenced are aware of the transactional nature of the exchange, that their data is being collected for profit, or the extensive long term social harms they are perpetuating through this medium at scale. Social media gave oxygen to the influencer economy as a form of marketing that leverages the popularity and reach of influential individuals to promote products and services.\cite{Freberg2011WhoPersonality} Notably, these marketing campaigns often involve digitally brushed images to present an idealised representation of reality.

\paragraph{}
This tendency to portray a perfect aesthetic online has raised concerns about authenticity and the fostering of unrealistic consumer expectations.\cite{Dwivedi2021SettingPropositions} Despite this, the influencer marketing industry has grown rapidly, with marketers increasingly recognising its potential for driving brand awareness and consumer engagement.

\section{Harms of Influencer Marketing}

The rise of generative AI technologies, which can autonomously create high-quality content, is altering the landscape of influencer marketing. Generative AI, such as Remini and Open AI’s Dall-E combined with ChaptGPT-4, can create authentic-looking images, videos, and written content, essentially becoming virtual influencers.\cite{Huang2018ArtificialService}

\paragraph{}
With this transition, the authenticity concern could take a new form: distinguishing real from AI-generated influencers and maintaining the authenticity of reality by human influencers.  Remini AI, a cutting-edge generative AI enterprise from Italy, is making waves across 'Black Twitter'. The firm offers various services, such as unblurring and sharpening images, noise reduction, restoring old photos, enlarging images, fixing colours, enhancing faces, and improving backgrounds. Leveraging these capabilities, Remini AI pledges to create diverse renditions of users' images from the user-provided images.
\paragraph{}
Remini is just one player in an increasingly crowded field of generative AI enterprises. What's captivating, though, is the emerging trend among Instagram influencers. They are using generative AI technologies like Remini to create virtually perfect images. These enhanced images amplify their brand and improve their visibility on Instagram's 'Explore' pages. The evolution of this industry will be influenced by how society and regulatory bodies respond to these advancements.

\subsection{Behaviours that Contribute to Harm
}
Certain behaviours contribute to potential harm in AI influencer marketing on platforms such as Twitter or Instagram. Influencers frequently disseminate compelling narratives, emphasising the competitive advantage of AI utilisation while neglecting potential risks. This technique primarily revolves around content creation and audience building, with little regard for potential consequences. 
\paragraph{}
On Instagram, transparency deficiencies are a prevalent issue in influencer marketing. Influencers often use AI enhancements for their photos without disclosing this manipulation, particularly when promoting beauty products. This can set unrealistic expectations and potentially jeopardise followers attempting to emulate these artificially enhanced images.\cite{McBride2019DigitalHealth} This failure to disclose, especially in the context of 'glow' products advocating skin lightening, can lead to harmful social outcomes.\cite{Fardouly2015SocialMood}
\paragraph{}
Moreover, influencers often endorse AI tools without adequately revealing their financial motivations, creating a false impression of impartiality.\cite{Abidin2016VisibilityInstagram} These undisclosed, financially-driven endorsements present serious ethical implications, as they may mislead consumers into adopting services or tools that they would not have chosen if privy to the underlying motivations. The question of who takes responsibility for ethical issues surrounding AI has been constantly debated across different sectors.\cite{Naik2022LegalResponsibility} However, in 2022, reality television star Kim Kardashian was penalised \$1.26 million for neglecting to reveal her collaboration with EthereumMax while promoting them to her extensive online following.\cite{SecuritiesandExchangeCommission2022SECSecurity} While laws and sanctions exist for the influence of investments and affiliate marketing without disclosure, laws have not kept up with the marketing of AI products or the encouragement of data sharing for profit.
\paragraph{}
Despite their enthusiastic endorsement, many influencers often need more comprehensive knowledge of the AI tools they promote. This could expose their followers to unseen risks associated with these tools, including privacy concerns.\cite{Hudders2023TheRecommendations} Thus, inadvertently promoting potentially hazardous tools and companies' varying data handling practices could put consumers at significant privacy risk. 
\paragraph{}
Apart from privacy concerns, significant cultural and societal detriments exist too. Although digital natives might understand these images to be enhanced or manipulated, digital immigrants may struggle to grasp this concept, leading to a distorted perception of products, individuals, and their capabilities.

\subsection{Driven by Capitalism
}
Capitalist motives primarily fuel AI-edited images. The entities behind these products are typically tech companies and digital marketers who recognise the commercial potential of visually enhanced content.\cite{Dwivedi2021SettingPropositions} By promoting 'perfected' visuals, these stakeholders can drive consumer engagement, stimulate desire, and ultimately encourage purchase behaviours. 
\paragraph{}
Technology firms monetise by selling AI-powered software and applications designed for image augmentation. These solutions employ advanced machine learning and deep learning models to transform visual content, empowering users to digitally modify their appearances.\cite{Schwartz2015TheMedia} While some argue that having an abundance of images representing diverse individuals can benefit society, it remains to be seen whether individuals are adequately informed and consenting to their images being used for further research. Often, such consent is buried within lengthy and complex terms of service agreements. What is known, is that followers of these influencers are not sharing in the reward of the profit generated from their data, resulting in more unpaid labour and exploitation.

\section{Ethical Data Collection, Responsible AI Development, and the Path Forward
}
Recent scholarly research\cite{Jaiswal2022RebootingCorporations}\cite{Wilson2019CreatingFuture} advocates for the simultaneous presence of workers and AI systems. This coexistence requires businesses to adopt a forward-looking approach to integrating AI, being mindful of their treatment of employees.\cite{Li2019CorporateTransformation} The objective is to create a harmonious coexistence between human workers and AI technologies in work environments, prioritising ethical concerns and the welfare of workers.\cite{Zirar2023WorkerAgenda} Acknowledging the historical burdens of colonialism and oppression, we must halt exploitative systems and transition toward adaptive solutions that benefit all of humanity and the planet. The status quo is unsustainable. If integrated and directed appropriately, AI can play a pivotal role in driving positive change. In an ideal scenario, workers and workplace AI complement each other by leveraging their respective strengths.\cite{Henkel2020HalfRegulation}\cite{Raisch2021ArtificialParadox} This coexistence allows workers to harness the precision, numerical capabilities, and pattern recognition of AI applications.\cite{Klotz2018HowCompetencies} By training AI to handle repetitive tasks accurately, workers can redirect their human expertise toward complex decision-making and critical analysis.\cite{Aoki2021TheExperiment}\cite{Shrestha2021AugmentingChallenges}\cite{Wilson2019CreatingFuture} The synergy between human intelligence and AI technologies fosters increased efficiency, productivity, and innovation in the workplace.\cite{Zirar2023WorkerAgenda}
\paragraph{Transparency and Accountability:}
Despite efforts to train AI systems to be explainable and transparent, a significant concern persists regarding AI solution providers and their willingness to disclose algorithmic details\cite{Davenport2018CanProblem} and to inform people when AI is being used, or when they are communicating with an AI and not a person. Relying on corporations to self-regulate is impractical. Global systems require comprehensive regulations to prevent exploitation. Companies that brand themselves as "ethical" cannot claim this status if they perpetuate exploitative practices, particularly by relying on labour from the Global South. To achieve sustainable development, the focus should shift away from prioritising corporate profit and fast-paced development, and instead, emphasise the respect and humane treatment of workers and prioritise environmental protection for future generations.

\paragraph{Addressing Current Issues in Data Collection and Labelling:
}
The issues associated with data collection and labelling, rooted in oppression and exploitation, are pervasive throughout industries. Companies operating in this field, including those mentioned in this paper, must be subject to regulation. It is crucial to enforce and protect the rights of workers and all individuals who share their data online. Many people remain unaware that their data is being collected and the potential dangers that come with it. This paper aims to raise awareness about the harms of such practices, where data is exploited by AI and the corporations behind it, and sometimes even misused by scammers.
\paragraph{Promoting Transparency in Data Collection:
}
Transparency in data collection is paramount. Individuals should be informed when their data is being collected, used, and shared. This necessitates clear communication from AI influencers when they collaborate with companies, and full disclosure of sponsored content in advertisements. People need to know when they are providing data, and they must understand the purpose behind its collection and usage.

\paragraph{Protecting Workers in Data Collection and Labelling:
}
Supposing that workers' engagement with workplace AI is aimed at compensating for AI's shortcomings,\cite{Wilson2019CreatingFuture} the AI that gains advantages from interacting with workers, rather than workers directly benefiting from these interactions. Such a situation may not align with the AI Principles of the Organisation for Economic Co-operation and Development (OECD), which emphasise that AI should primarily benefit people.\cite{OECD2021StatePolicies}\cite{Zirar2023WorkerAgenda}  Workers involved in data collection and labelling should receive fair compensation in addition to fair treatment. Additionally, they should have a comprehensive understanding of the work they are undertaking and the implications of the data they handle. Their rights as employees must be safeguarded, and measures should be in place to protect them from exploitative practices. A truly empowering approach in workplaces involves fostering a symbiotic relationship between workers and intelligent systems to overcome limitations.\cite{Wilson2019CreatingFuture} However, it is essential for companies to prioritise workers' rights by ensuring that the workforce possesses the necessary skills for this symbiosis.\cite{Sousa2018SustainableAge}

\section{Conclusion}
Unethical data collection and labelling practices pose significant risks to individuals and society at large. To progress ethically, we must prioritise transparency, accountability, and fair treatment of workers. By regulating companies and promoting responsible AI development, we can pave the way for a sustainable future that benefits everyone. It is time for a collective effort to address these challenges and steer AI technology toward a positive and inclusive future. By upholding workers' rights and providing them with the required skills, we can create a collaborative and mutually beneficial environment where both humans and AI technologies thrive. 

While this short review has shed some light on pertinent issues around ethical considerations for AI-driven digital value networks, an in-depth investigation is warranted in future research to fully understand the implications and potential regulation of the practices outlined in this piece.

\begin{ack}
We acknowledge the Berkman Klein Center for Internet and Society at Harvard for hosting the AI and Agency in the Global South working group which brought these authors together.
\end{ack}

\printbibliography

\end{document}